\documentclass[twocolumn,amsmath,amssymb,superscriptaddress,longbibliography]{revtex4-1}

\usepackage{color}
\usepackage{graphicx}
\usepackage{amssymb}
\usepackage{amsmath}
\newcommand{\ud}{\mathrm{d}}

\newcommand{\sech}{\mathrm{sech}}
\newcommand{\sgn}[1]{\mathrm{sgn} #1}

\graphicspath{ {./figures/} {./}}

\begin{document}

\title{Geometric origin of rogue solitons in optical fibres}

\author{Andrea Armaroli}
\email{andrea.armaroli@mpl.mpg.de}
\affiliation{Max Planck Research Group `Nonlinear Photonic Nanostructures' \\Max Planck Institute for the Science of Light, G{\"u}nther-Scharowsky-Str.~1/Bau 24
91058 Erlangen, Germany}
\author{Claudio Conti}
\affiliation{Institute for Complex Systems (ISC-CNR) and University of Rome ``La Sapienza'', Department of Physics, Piazzale Aldo Moro 5
00185, Rome, Italy}
\author{Fabio Biancalana}
\affiliation{Max Planck Research Group `Nonlinear Photonic Nanostructures' \\Max Planck Institute for the Science of Light, G{\"u}nther-Scharowsky-Str.~1/Bau 24
91058 Erlangen, Germany}
\affiliation{School of Engineering and Physical Sciences, Heriot-Watt University, EH14 4AS Edinburgh, United Kingdom}
\date{\today}

\maketitle
{\bf Non-deterministic giant waves, denoted as rogue, killer, monster or freak waves,  have been reported in many different branches of physics.
Their origin is however still unknown:
despite the massive numerical and experimental evidence, the ultimate reason for their spontaneous formation has not been identified yet. 
Here we show that rogue waves in optical fibres actually result from a complex dynamic process
very similar to well known mechanisms such as glass transitions and protein folding.
We describe how the interaction among optical solitons 
produces an energy landscape in a highly-dimensional parameter space with multiple quasi-equilibrium points.
These configurations have the same statistical distribution of the observed rogue events and 
are explored during the light dynamics due to soliton collisions, with inelastic mechanisms enhancing the process. 
Slightly different initial conditions lead to very different dynamics in this complex geometry; a rogue soliton turns out to stem from one particular deep quasi-equilibrium point of the energy landscape in which the system may be transiently trapped during evolution.
This explanation will prove fruitful to the wide community interested in freak waves. }

Observations of non-deterministic giant wave events have been reported in many different fields, from oceanic dynamics \cite{Hopkin2004,Garett2009}, to financial markets \cite{Zhen-Ya2010} and light propagation in optical fibres \cite{Solli2007,Jalali2010,Solli2010,DeVore2013}.
The origin of these phenomena, denoted as rogue, killer, monster or freak waves, is still unknown. Since the first report of optical rogue solitons (RSs) in the supercontinuum generation (SCG) occurring in optical fibres \cite{Solli2007}, a great effort has been put in characterising these light pulses with extremely large amplitudes.   
Despite the massive numerical and experimental evidence \cite{Dudley2008,Dudley2009,Genty2010,Jalali2010,Solli2010,Sorensen2012,Akhmediev2013,Vergeles2011,Oppo2013}, the ultimate reason for their spontaneous formation has yet to be identified. 
While it was suggested that RSs are exact solutions of the nonlinear Schr{\"o}dinger equation (NLS) such as Akhmediev's breather \cite{Akhmediev1986a,Akhmediev2009,Dudley2009} or the Peregrine soliton \cite{Kibler2010}, or rather solutions of a more general model \cite{Ankiewicz2012,Bandelow2013,Baronio2012}, there are strong indications that they originate from a combination of higher-order effects such as Raman scattering, self-steepening and higher-order dispersions \cite{Dudley2008,Mussot2009}. 
Exact solutions are pivotal to explain the nonlinear stages of modulation instability (MI), which is observed under continuous wave excitation, but their properties
(amplitude, finite background, periodic nature) are in contrast to the arbitrary strong pulses which are detected when exciting the optical fibre with ultrashort pulses.
More importantly this class of solutions does not give any information about the statistical distribution of high intensity peaks, which like many extreme natural and social phenomena, exhibits a heavy-tail behaviour \cite{Solli2007,Erkintalo2010,Sorensen2012}.

At variance with previous studies, we treat the nonlinear light propagation in an optical fibre as a {\em dynamical system out of equilibrium}. 
In a complex medium such as a molecular glass or soft-colloidal matter, the local interaction between adjacent particles forms a global potential, the \emph{energy landscape} \cite{Goldstein1969,Stillinger1984,Sastry1998,Debenedetti2001,Conti2005}.  
The landscape is formed by many local minima and is a concept used in different branches of physics to describe, for example, the supercooling phase transition from a liquid to a glass.

We prove that our system explores the energy landscape generated by the weak interactions between neighbouring NLS solitons, while the dynamics is accelerated by the Raman effect, which leads to a continuous growth of an {\em equivalent temperature}.
Thus a RS turns out to have its origin in one of the quasi-equilibrium points of the energy landscape, which corresponds to a deep minimum in the interaction energy. This minimum also corresponds to collisions between high energy pulses emerging during the SCG process.

Finally, these quasi-equilibrium points in the landscape exhibit the same heavy-tail statistical distribution of the peaks detected in SCG experiments. This explanation will prove to be extremely useful to all those communities dealing with the study of freak waves
\cite{Bludov2009,Garnier2010,Ruderman2010,Stenflo2010,Onorato2010,Arecchi2011,Onorato2013}.

\section{The energy landscape}

\begin{figure*}
	\centering
		\includegraphics[width=0.8\textwidth]{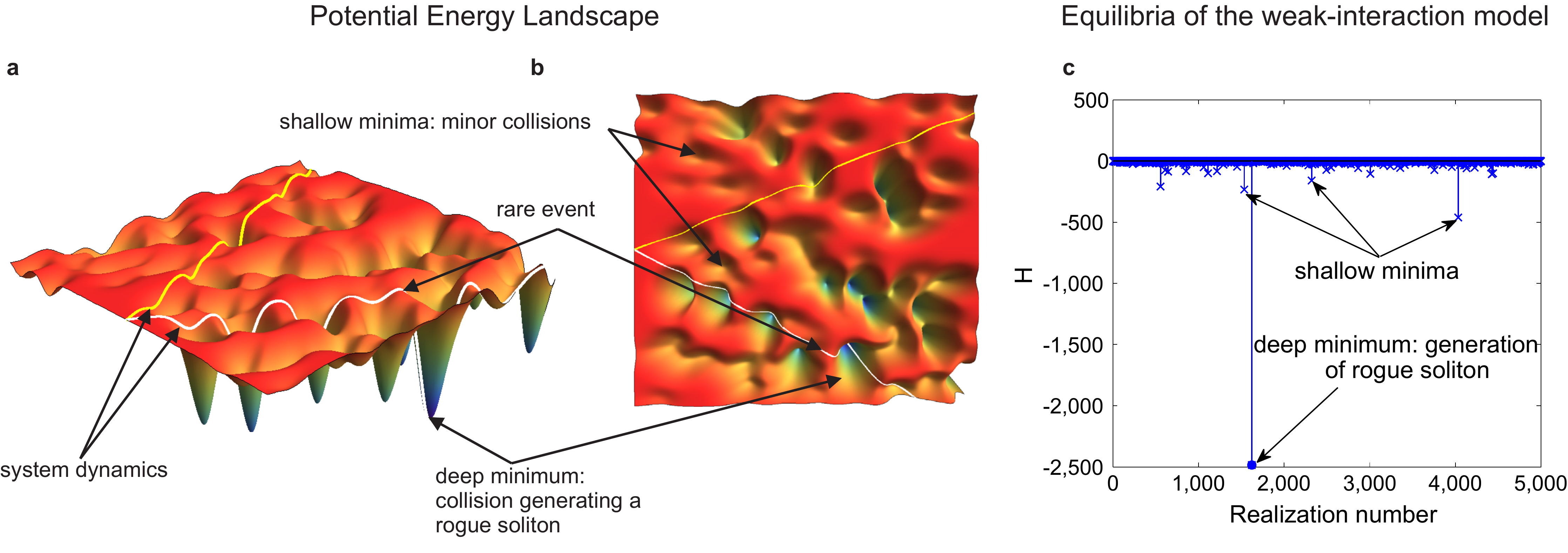}
	\caption{{\bf Potential energy landscape and the emergence of rogue solitons in optical fibres}
(a-b) A pictorial representation of the energy landscape of a complex system represented as a two-dimensional surface: we observe many local minima with various depth and location, separated by saddle points. The dynamics is strongly influenced by the initial conditions: we highlight two different trajectories, one dropping into a deep minimum of the potential energy (white line) and 
one visiting many shallow minima (yellow line).  This explains why slightly different noise configurations around the input pulse generate completely different dynamics, leading only rarely to rogue solitons. (c) We compare this simplified image with the interaction of many 
solitons. It is characterised by the evolution in a multidimensional complex topology with many minima and saddles. Rare events, due to collisions or higher order effects (as dispersion and Raman gain), trigger the system in regions with lower energy, which is in our case the Hamiltonian of the NLS. We represent the energy landscape generated by soliton interaction {by} the different values of $H$ at different numerically-computed equilibria. The horizontal axis enumerates the solutions obtained from different initial random conditions. The solution marked with a full circle is the exceptional solution, which we show in Fig.~\ref{fig:Nsol40}.
	\label{fig:PEL}}
\end{figure*}

The energy landscape can be built by starting from the simplest model of soliton propagation in fibre optics: the dimensionless NLS equation in anomalous dispersion {\cite{Agrawal2012}}
\begin{equation}
	i\frac{\partial{u}}{\partial z} + \frac{1}{2}\frac{\partial^2 u}{\partial t^2} +|u|^2 u=0,
\label{eq:NLS}
\end{equation}
where $u(z,t)$ is the normalised electric field envelope of the pulse, $z$ is the longitudinal coordinate along the fibre and $t$ is time.
We study a regime, commonly observed in experiments and numerical simulations, in which the highly
nonlinear dynamics can be described as a system of $N$ weakly-interacting solitons.
The conventional strategy consists in applying  a variational approach and assuming for the solution of Eq.~\eqref{eq:NLS} the following \emph{Ansatz},
\begin{equation}
u(z,t) = \sum_{k=1}^N{u_k(z,t)}
\end{equation} 
with
\begin{equation}
	u_k(z,t) = 2\nu_k\,\sech{\left[2\nu_k (t-\xi_k)\right]}e^{i2\mu_k(t-\xi_k)+ i\delta_k}
\label{eq:Nsol}
\end{equation}
where $\nu_k$, $\mu_k$, $\xi_k$ and $\delta_k$ represent the amplitude, speed, temporal delay and phase of each soliton, labelled by an index $k=1,\ldots,N$.
It is well established that one can model the nearest-neighbour interactions of NLS solitons as a system of ordinary differential equations (ODEs) for the $4N$ parameters of Eq.~\eqref{eq:Nsol}
\cite{Uzunov1996,Gerdjikov1996,Anderson1985,Anderson1986,Arnold1993,Karpman1981a, Agrawal2012,Turitsyn2012}. We denote this model (detailed in Methods) as the weak-interaction model (WIM). 

The WIM can be reduced to a mechanical system under the hypothesis of nearly equal amplitudes $\nu_k\approx \nu$. However we follow an approach similar to Ref.~\cite{Conti2005}, where the appearance and stabilisation of filaments in a nonlocal medium was explained as a {\em complex phase transition of light}, and we implement a numerical search for the fixed points (stable or unstable equilibria, see Methods) of the WIM, i.e. those points for which $\dot\nu_k=\dot\mu_k=\dot\xi_k=\dot\delta_k=0$.

The basic idea here is to map the highly nonlinear dynamics of rogue wave generation to a \emph{complex dynamical topography in a highly-dimensional system}, as it is performed in many other branches of physics, like supercooled liquids, chemical reactions and protein folding \cite{Goldstein1969,Stillinger1984,Sastry1998}.
In those systems it is convenient to represent the interactions as a potential energy landscape, as sketched in Fig.~\ref{fig:PEL}(a,b), for a two-dimensional system. This landscape exhibits many minima of different depths, and saddle points, along which the system can escape and jump randomly from one configuration to another. 
In a multidimensional setting the central parameter is the \emph{saddle order}, which quantifies the ability of the system to escape from a specific minimum configuration. 

The WIM corresponds to a $4N$-dimensional \emph{phase space} and we now discuss the properties of its equilibrium points in terms of saddle order and their relation to minima of an equivalent global energy of the system. 
We discover that the statistical distribution of these points have exactly the same features of the rare-events due to rogue wave generations observed in experiments.

The numerical procedure adopted here, and detailed in Methods, allows us  to find, for a fixed $N$, a large variety of fixed points, amongst which we identify
{\em exceptional solutions}, which are characterised by a large amplitude soliton $\nu_k$, and that correspond to the rogue-soliton-generating collisions. Moreover we find many other solutions, which comprise in general smaller   multiple unevenly-spaced intensity peaks. 

\begin{figure*}
	\centering
		\includegraphics[width=0.8\textwidth]{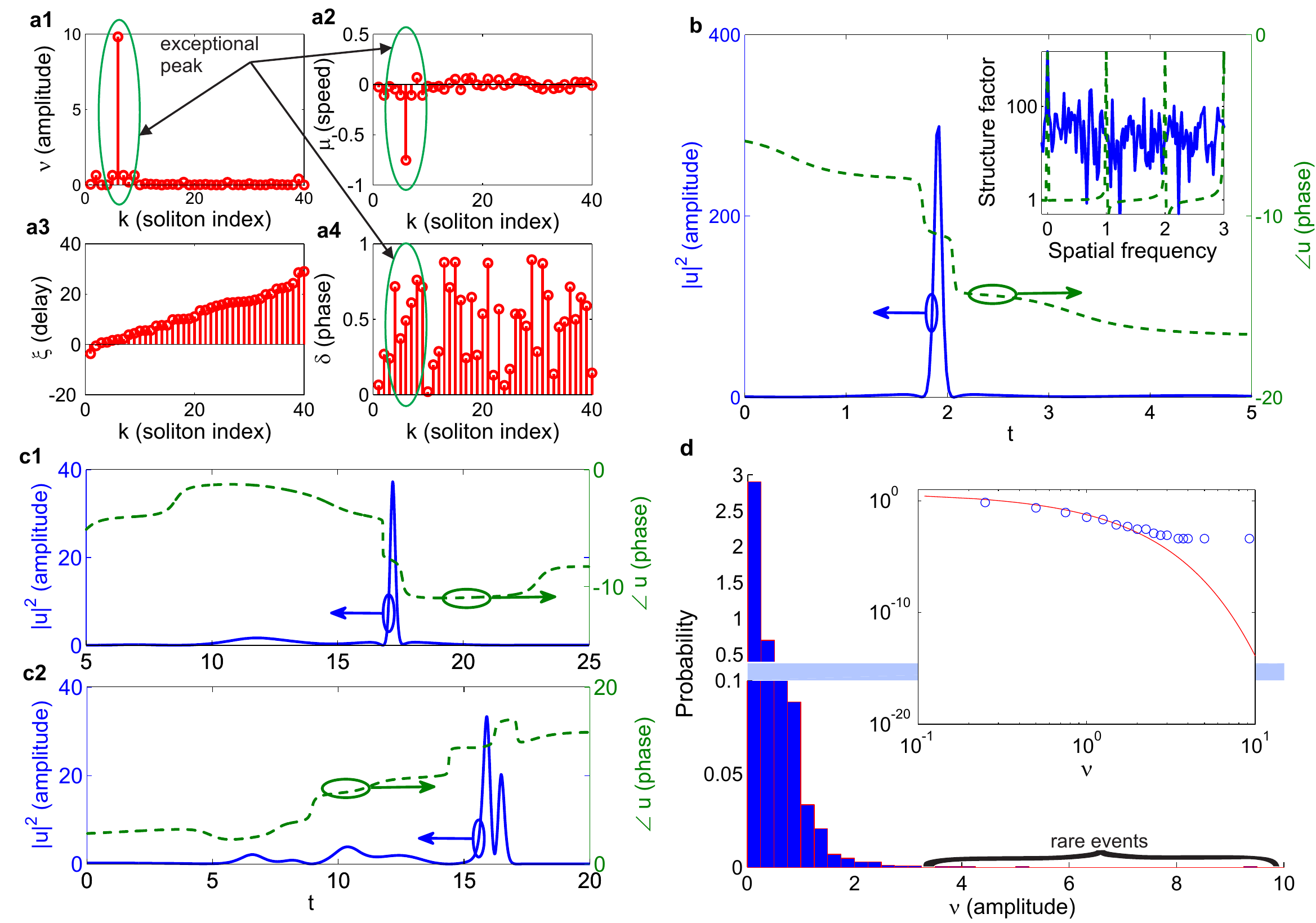}
	\caption{{\bf Equilibria of the weak interaction of NLS solitons (N = 40) and their statistical properties.}  (a-b) Characterization of an exceptional solution with anomalously large peak: (a) fixed point of WIM, sorted according to the position $\xi_k$; (b) squared amplitude and phase of the reconstructed field  which shows a three peak structure, and an apparent sign change across the pulse (the phase exhibits two subsequent jumps); in the inset we  compare the structure factor of the present solution (blue solid line) with a crystal the lattice constant of which is $\Delta\!\xi_\mathrm{ave}$ (green dashed line). In contrast to (b), we show in (c) two different equilibrium solutions of the WIM which are much less intense and present a multiple peak structure: in (c1) the peak is one order of magnitude smaller than in panel (b) and a trailing pulse appears, while (c2) {exhibits} two peaks. These solutions correspond to the shallower minima in Fig.~\ref{fig:PEL}(c). Finally in (d) the statistical analysis of the solutions of WIM is presented: (d) statistical analysis of the ensemble of the WIM solutions: histogram of the amplitudes exceeding the $Q=0.95$ quantile of the distribution. In the inset the Weibull fit of the distribution in double logarithmic scale ($N=40$).}
	\label{fig:Nsol40}
\end{figure*}

In Fig.~\ref{fig:Nsol40}(a-b) we show an example of an exceptional solution.
We observe a typical pattern: a large amplitude soliton (a1) is closely surrounded by two lower amplitude solitons [see (a3)],
and many others with smaller peak amplitudes. The velocity of the main soliton is much larger than that of its neighbours (a2); phases are ordered across the triplet. We observe thus a solution composed by a large peak surrounded by smaller oscillations. This is confirmed by the plot of the squared modulus and phase, Fig.~\ref{fig:Nsol40}(b). The RS solution appears as a dominant peak superimposed with a smaller one with a different speed. This gives rise to oscillations and phase jumps across the profile. Any exceptional solution is actually composed by many solitons, but these are arranged in a non-periodic,
disordered fashion (a {\em glass} of solitons, as defined in the introduction). Indeed, we show in the inset of  Fig.~\ref{fig:Nsol40}(b) the {\em structure factor} (defined in Methods) of our specific exceptional solution and compare it to that of a regular lattice with a lattice constant equal to the average separation occurring in our solution.  In contrast with regularly spaced equal peaks of an ordered arrangement, the plot exhibits irregularly spaced peaks, typical of glassy systems.

The WIM solutions of the kind shown in Fig. \ref{fig:Nsol40}(b) seem to qualitatively resemble the Peregrine soliton \cite{Kibler2010} or a slice of an Akhmediev breather \cite{Akhmediev1986a,Dudley2009}. This is though misleading, because in our case we do not have an infinite background, but a sum of pulses of different amplitudes, which decay exponentially. The ratio of amplitudes is not fixed and solutions with two peaks are also found. Instead the Peregrine soliton and similar solutions of higher-order models \cite{Bandelow2013} exhibit a fixed ratio of the peak versus the background. In order to further support the energy landscape hypothesis we report in Fig.~\ref{fig:Nsol40}(c1,c2) two different equilibrium solutions: the peak intensity is one order of magnitude smaller than in Fig. \ref{fig:Nsol40}(b), and the field profile can have single or multiple peaks.

We remark two fundamental facts:  first the considered solutions are unstable fixed point of the WIM, with saddle order (i.e.~number of eigenvalues with positive real part, see Ref. \cite{Conti2005}) $N_s \approx 2N$. Higher-order effects in optical fibres permit to explore a broad range of irregular solutions and the saddle order hints at the probability of falling into such a solution.
Second the exceptional solution corresponds to a minimum value (among the numerically found solutions) of the integrated Hamiltonian density of the NLS, see Methods.

According to this argument, in Fig.~\ref{fig:PEL}(c) we show a representation of the energy landscape in the form of an enumeration of the values of $H$ obtained numerically from each different initial guess. Each point in Fig.~\ref{fig:PEL}(c) is a quasi-equilibrium configuration. We notice that most solutions correspond to small or vanishing quantities, many solutions with $H>-10^{3}$ are found and the exceptional solution of Fig.~\ref{fig:Nsol40}(b) clearly emerges. Shallower minima correspond to solutions such as those shown in Fig.~\ref{fig:Nsol40}(c), with smaller peak amplitudes and a multi-peaked structure.

The statistical analysis of the ensemble of solutions obtained by the numerical solver is shown in Fig.~\ref{fig:Nsol40}(d). We select the values of $\nu$ exceeding the $Q=0.95$ quantile and show that it exhibits a heavy-tail behaviour. To further confirm this we fit the data to a Weibull distribution, universally used in the study of rogue wave events in optical fibres: this probability density function is a prototypical heavy-tail distribution which occurs in the description of many extreme phenomena and was found earlier in the distribution of rogue solitons \cite{Dudley2008}. The same can be done for $-H>0$ (not shown). We obtain again a heavy-tail behaviour as expected. 
We remark that also the statistical moments are similar for the distributions of $\nu$ and $H$. Specifically we use two parameters which summarise the deviation of the distribution from the Gaussian case: the skewness (3rd moment) $\bar\gamma$ and the kurtosis (4th moment) $\kappa$. We compute $\bar\gamma\approx 4$ and $\kappa \approx 20$, so that their product is $\bar\gamma\kappa\gg 10$: this condition characterises a  heavy-tail statistics and corresponds exactly to what was found in previous rogue soliton experiments \cite{Sorensen2012}. This gives us further confidence that the WIM model is an appropriate description of the physics behind rogue soliton formation in fibres.
In summary we are able to draw for the first time a map of the {\em topography} of the energy landscape for the soliton interactions, which turns out to comprise  many disordered solutions, among which a strongly peaked one emerges.

\section{Simulations}
Next we simulate light propagation in an optical fibre and compare the characteristic features of rogue wave formation with the exceptional solutions of the WIM. We consider the well-tested and universally used generalised nonlinear Schr{\"o}dinger equation (GNLS) and performed accurate simulations with parameters similar to \cite{Dudley2008}, see Methods for details.
\begin{figure*}%
\includegraphics[width=0.75\textwidth]{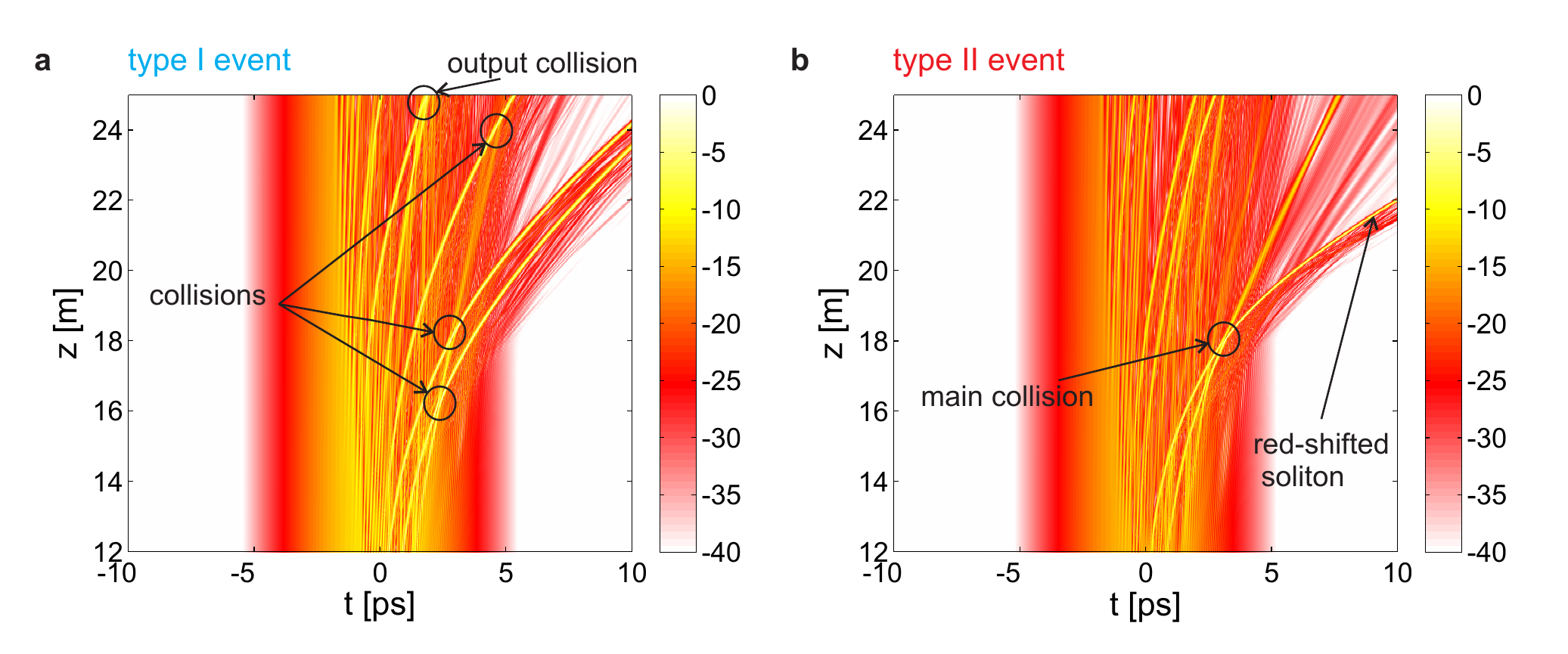}
\caption{{\bf Evolution of the temporal profile along the propagation direction:  at $Z=25$ m} (a) corresponds to the maximum peak (type I rogue soliton), (b) corresponds to the most red-shifted output (type II rogue soliton). In (a) we observe a sequence of collisions ending in an event at the fibre output. In (b) we identify the collision event which leads to the emergence of the most red-shifted soliton. 	}
\label{fig:GNLSresults}%
\end{figure*}

After a propagation length $L = 25$ m, we identify the two most important scenarios:  among all the realisations of noise (i) the output exhibits the strongest peak---we define it as \emph{type I RS}---and (ii) the output contains the most red-shifted soliton---we denote it as \emph{type II RS}. 
We report these two situations in Fig.~\ref{fig:GNLSresults}(a) and (b) respectively. It is a well-established fact \cite{Erkintalo2010} that the strongest peak is not necessarily the most red-shifted, due to the role played by collisions in locally enhancing intensity.
As described in the following, we find that the two scenarios correspond to two  different kinetic processes in the energy landscape.

After the initial stage of MI and pulse train generation, we observe between 16 m and 18 m that collisions start to occur. 
In the first scenario, see Fig.~\ref{fig:GNLSresults}(a), they do not result in a dramatic splitting of the spectrum into two parts (corresponding to the red-shifted soliton and the main part of the pulse train), while many subsequent events lead to the observation of a peak at the fibre output, which is itself a collision. The second scenario, see Fig.~\ref{fig:GNLSresults}(b), entails a more dramatic early event, which transfers a large amount of energy to a single soliton and leads to the appearance of a strongly red-shifted spectral part. We show that these sequences of inelastic collisions correspond to hopping from one 
metastable equilibrium point to another in the energy landscape. We find that it is crucial to follow the evolution of type I and type II rogue soliton generation mechanisms beyond $L=25$ m and extend our simulations up to $L'= 40$ m.
 
\begin{figure*}%
\includegraphics[width=0.9\textwidth]{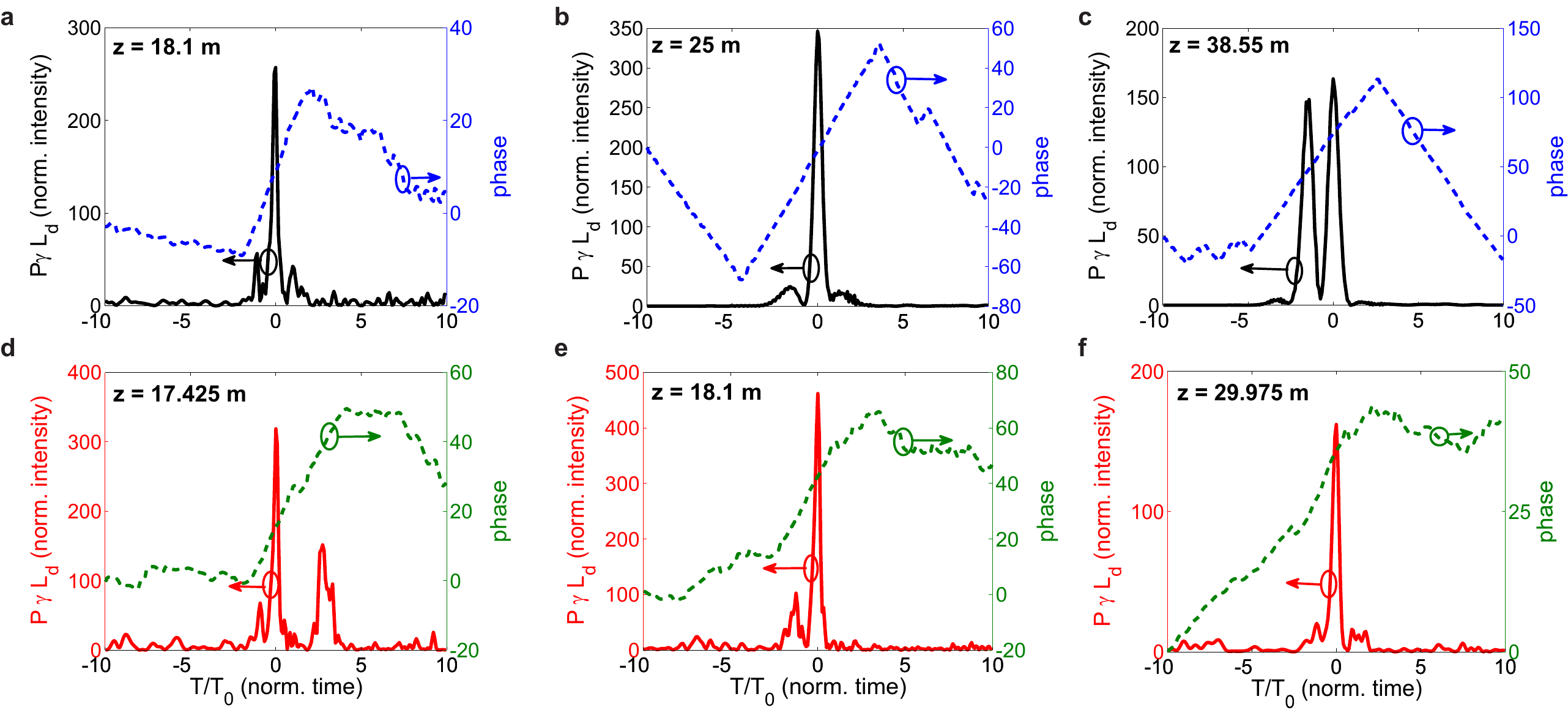}
\caption{{\bf Temporal profile at collision points.} The time variable is centred and scaled by $T_{M\!I}$ and we included events occurring after $Z=25$ m. (a-c) shows three collision events for the type I event: while a collision occurs near $Z\approx18$ m, see Fig.~\ref{fig:GNLSresults}(a), a major event occurs at $Z\approx25$ m. In (d-f) we consider type II scenario: in this case the collision at $Z \approx 18$ m is the main event and the amplitude peak is much stronger than at any other propagation stage. The amplitude and phase waveform are in qualitative agreement between the  analytical model and the simulation results, see Fig.~\ref{fig:Nsol40}(b-c). 
}%
\label{fig:collisions}%
\end{figure*}

In Fig.~\ref{fig:collisions} we show the details of three collision events for each of the scenarios shown in Fig.~\ref{fig:GNLSresults};
we rescale time and field amplitude (see Methods) in order to compare easily with the solution of the WIM in Fig.~\ref{fig:Nsol40}.

With reference to a type I event shown in Fig.~\ref{fig:GNLSresults}(a), Fig.~\ref{fig:collisions}(a, b) show events occurring at  $Z\approx18$ m (the first collision after the pulse train formation), and $Z\approx25$ m, which leads to the large output peak.
Fig.~\ref{fig:collisions}(c) shows a further collision at a longer propagation distance $Z\approx39$ m with two main peaks, like the one of Fig.~\ref{fig:Nsol40}(c2).

With reference to Fig.~\ref{fig:GNLSresults}(b), type II event,
Fig.~\ref{fig:collisions}(d, e) deal with the generation of the most red-shifted soliton, which stems from a collision at $Z\approx18$~m (e) where the power level reaches over $2\times10^{3}$ W [or about 500 in normalised units], about twice as in the case of Fig.~\ref{fig:collisions}(b). 
This is indeed the main event in SCG (and the most studied after the original observations of Ref.~\cite{Solli2007}), as can be noted by comparing it to a collision occurring slightly  before, at $Z\approx17.5$~m [Fig.~\ref{fig:collisions}(d)],  and one at a much further distance $Z\approx30$~m [Fig.~\ref{fig:collisions}(f)]. 

We stress the similarity in amplitude and temporal width of the solutions of the WIM, Fig.~\ref{fig:Nsol40}(b), to the temporal profile in Fig.~\ref{fig:collisions}(b, e). A large peak of amplitude $|u|^2\approx 300$ and width $\Delta t\approx1$  is surrounded by small nearly symmetric oscillations. The minor events occurring before Fig.~\ref{fig:collisions}(a, d) and after the main collision Fig.~\ref{fig:collisions}(c, f) strongly resemble the solutions with smaller peaks presented in Fig.~\ref{fig:Nsol40}(c), compare panel (f) to the double-peak solution of panel Fig.~\ref{fig:Nsol40}(c2).
The phase profile is strongly distorted by the Raman acceleration, i.e.~a $Z$-dependent  phase slope proportional to the fourth power of the soliton amplitude. Apart from that and possibly an overall slope (due to the choice of the reference frame), the same characteristic jump sequence of Fig.~\ref{fig:Nsol40}(b-c) may be recovered. 

We thus found that subsequent collisions resemble more and more the solution of the WIM, up to a major collision event Fig.~\ref{fig:collisions} (b, d). The collision at $Z\approx18$ [Fig.~\ref{fig:collisions}(d)], type II RS generation, exhibits a much more energetic event than what we observe in panel (b) in type I RS generation. 
Then minor interactions occur and the rogue soliton freely propagates, giving rise to the signature red-shifted part of the spectrum first observed experimentally in Ref. \cite{Solli2007}.
In other words, in type II rogue soliton, a collision triggers the system into the exceptional point of the landscape, and after that  the system is nearly quenched in its dynamics. The dynamics of type I events is apparently more gradual due to a sequence of minor collisions.

\begin{figure}%
\includegraphics[width=0.45\textwidth]{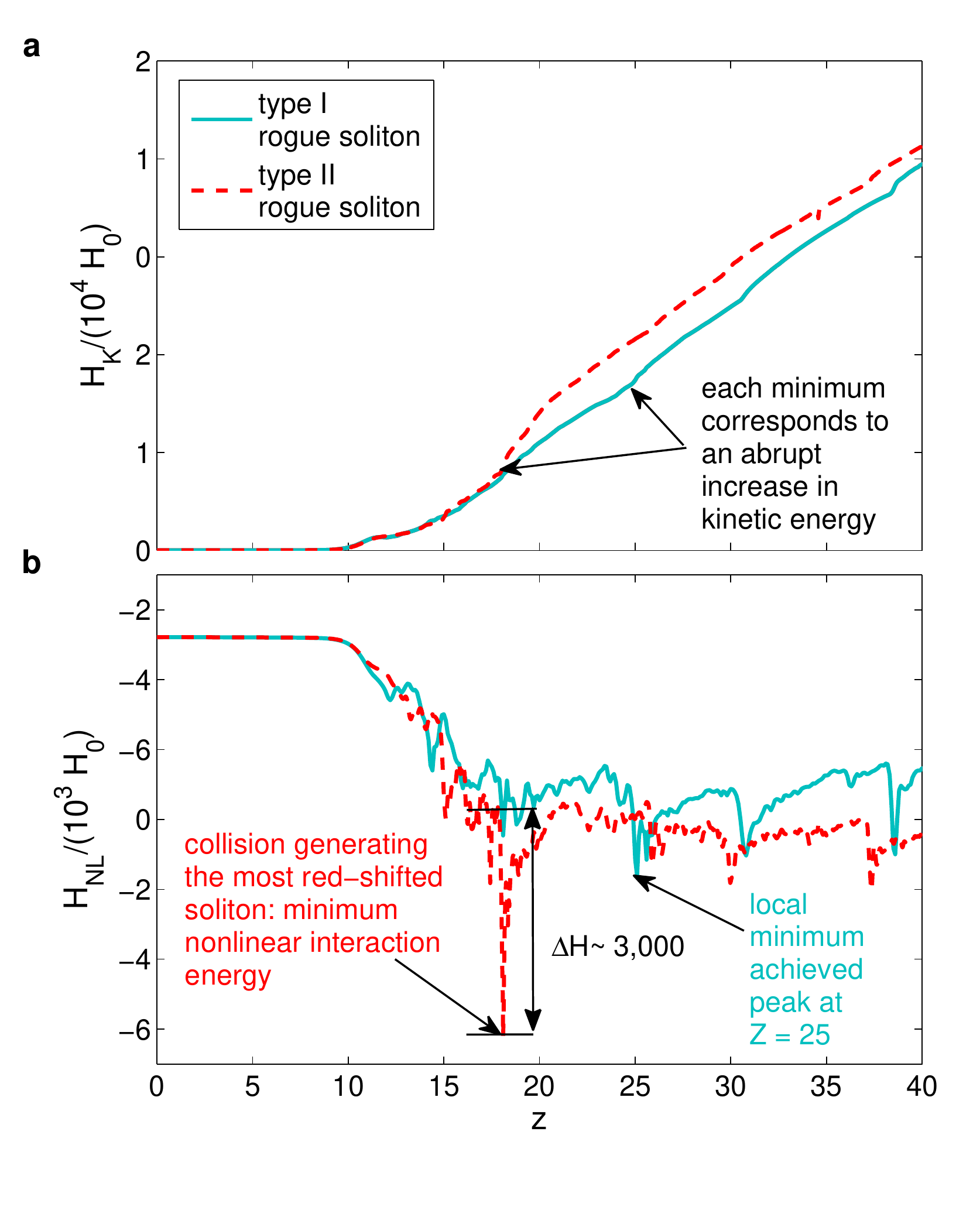}
\caption{{\bf  Hamiltonian evolution $H(z)$}. We separate the kinetic (linear) $H_K$ (a) and nonlinear part $H_{N\!L}$ (b) (see Methods). 
We report the two scenarios described in the text and shown in Fig.~\ref{fig:GNLSresults}: the solid blue line corresponds type I rogue soliton generation, the dashed red line to type II mechanism. In order to thoroughly characterise the evolution of the two scenarios, we show the evolution up to $L'=40$~m and normalise these quantities according to Methods in order to compare $H_{N\!L}$ with the energy landscape of Fig.~\ref{fig:PEL}(b). We notice that $H_{N\!L}$ exhibits many local minima, which correspond to step increase in $H_K$ and this latter tends to settle on a linear growth. Importantly each minimum corresponds to a soliton collision. The arrows highlight the main events at $Z\approx18$ m and $Z\approx25$ m, for the two scenarios. The minima correspond to an increase in $H_K$, which then settles on a steady linear growth. The deeper minimum in type II scenario leads to a sudden increase to larger values of $H_K$, while type I is characterised by smaller steps. 
 }\label{fig:hamiltonian}%
\end{figure}

The last decisive observation of our work is that after splitting the Hamiltonian into two parts, a kinetic $H_K$ (due to group velocity dispersion) and a potential $H_{N\!L}$ (originating from nonlinear interaction) part, see Methods, we notice that {\em the collision events correspond to minima of the potential part of the Hamiltonian} $H_{N\!L}(z)$.  
This is clearly visible in Fig.~\ref{fig:hamiltonian}. 
In order to compare the simulation results with the energy landscape of Fig.~\ref{fig:PEL}(c), we normalise the $y$-axis according to Methods.
While the kinetic part in Fig.~\ref{fig:hamiltonian}(a) plays the role of an  \emph{effective temperature} which grows mainly because of the Raman acceleration, the potential part $H_{N\!L}(z)$  exhibits a sequence of minima.
In this regard, the considered two scenarios in  Fig.~\ref{fig:GNLSresults} are fundamentally different: in the first, Fig.~\ref{fig:GNLSresults}(a),  the deepest minimum occurs at $Z\approx25$ m, after exploring shallower minima; the second, Fig.~\ref{fig:GNLSresults}(b), exhibits a much deeper minimum at $z\approx18$ m, and minor events follow. This is consistent with the collision sequence described above. Moreover the depth of the first local minimum in the type II scenario is of the same  order of $H$ as obtained for the exceptional solutions of the WIM.

We notice that contrary to the expected quadratic growth (see Methods), $H_K$ shows only a linear increase: the inelastic collisions among solitons behave as an {\em equivalent friction}, which puts an upper limit to the speed of the 'particles' in the system.

To summarise, it is possible to observe a collision event near the fibre output:  this corresponds to a regime characterised by jumps from one of the many shallow minima of the energy landscape to another. 
On the contrary, in the presence of a large-amplitude red-shifted soliton, an early collision lets the system fall inside a very deep minimum in the energy landscape. 
Upon propagation, the Raman effect leads to the growth of an effective temperature of the system, i.e., an increase of the kinetic energy, and inhibits the stabilisation in one equilibrium points.

We have also explored the effect of higher-order dispersion in the absence of the Raman effect. The convective instabilities \cite{Mussot2009} are undoubtedly an important ingredient for rogue solitons to be excited. Anyway the collisions in this case correspond to much shallower minima of $H_{N\!L}$ (not shown), so that the appearance of extremely strong peaks is substantially hampered.

Rogue solitons appear in the supercontinuum generation via successive collisions, which in turn are minima of the energy landscape that can be described by the WIM. 
We explored a scenario of slow RS formation, type I, which is explained by the hopping between many different equilibria of our energy landscape and a scenario of quick rogue soliton emergence, type II, in which the system sinks early into a deep minimum. In essence, rogue soliton formation is a rare event since slightly different initial configurations for the input noise can lead to very different trajectories in the phase space. When such trajectories intersect deep minima of the energy landscape, collisions occur that form strongly peak rogues solitons. The exceptional solution of the WIM represent an extremely rare collision event, while other minor equilibria correspond to less dramatic interactions. 
This {\em geometric} explanation finally sheds light on the previous---and quite mysterious---numerical and experimental observation that very similar noise configurations lead to completely different pulse dynamics. We conclude that the topography of the energy landscape is thus crucial to explain the formation of RSs and to classify their behaviour in terms of collision profiles and statistical properties. Understanding and controlling the energy landscape will lead to a better control of the rogue wave formation. To the best of our knowledge, this result represents the first actual {\em explanation} of the RSs formation in optics. The generality and universality of the concept will prove extremely useful to all those communities dealing with the study of freak waves.

\section{Methods}
\subsection{Weak interaction model}
In order to derive the weak interaction model, we start from Eq.~\eqref{eq:NLS} and use the following \emph{Ansatz} composed by $N$ weakly interacting NLS solitons,
and model their pair-wise interaction   as \cite{Uzunov1996,Gerdjikov1996,Anderson1985,Anderson1986,Arnold1993,Karpman1981a, Agrawal2012}
\begin{equation}	
\begin{aligned}
		\dot{\nu}_k &= 16 \nu_k^2 \left(S_{k,k-1} - S_{k,k+1}\right)\\
		\dot{\mu}_k &= -16 \nu_k^2 \left(C_{k,k-1} - C_{k,k+1}\right)\\
		\dot{\xi}_k &= 2\mu_k - 4 \left(S_{k,k-1} - S_{k,k+1}\right)\\
		\dot{\delta}_k &= 2( \nu_k^2 + \mu_k^2) - 8\mu_k \left(S_{k,k-1} + S_{k,k+1}\right) \\
		&+24\nu_k\left(C_{k,k-1} + C_{k,k+1}\right)
\label{eq:WIM}
\end{aligned}
\end{equation}
where {the dot denotes the derivative with respect to $z$}
and  
\begin{equation}
\begin{aligned}
S_{k,n} &= e^{|\beta_{kn}|}\nu_n \sin{s_{kn}\phi_{kn}}\\
C_{k,n} &= e^{|\beta_{kn}|}\nu_n \cos{\phi_{kn}}\\
\beta_{kn} &= 2\nu_k \left(\xi_k-\xi_n\right)\\
\phi_{kn} &= \delta_k - \delta_n - 2\mu_n\left(\xi_k-\xi_n\right)
\label{eq:WIMquant}
\end{aligned}
\end{equation}
and $s_{kn} = \sgn{\left[\beta_{kn}\right]}$.

\subsection{Numerical calculation of fixed points}
In order to find the fixed point of this system, we set $N$ and compute numerically the fixed points starting from $N_\mathrm{iter}=5,000$ randomly distributed initial conditions: amplitudes $\nu_k$ are log-normally distributed, velocity $\mu_k$ and positions $\xi_k$ are normally distributed, and phases $\delta_k$ follow a uniform distribution in $[0,2\pi]$.

Matlab\textsuperscript{\textregistered}  optimisation toolbox is employed, specifically the nonlinear solver \texttt{fsolve} with the Levenberg-Marquardt algorithm. Convergence is reliably achieved up to $N=40$. 

\subsection{Structure factor}
The structure factor is defined by $S(q) = \mathcal{F}\left[\sum_{n=1}^{N}\delta(\xi-\xi_n)\right](q)$ where $\mathcal{F}[\cdot]$ denotes the Fourier transform and we consider the single realisation represented by our special solution and compare to perfect crystal
$\Delta\!\xi_\mathrm{ave} \equiv \sum_{k=2}^N{\xi_k-\xi_{k-1}}$.

\subsection{On the splitting of NLS Hamiltonian density}

It is well known that the NLS  \eqref{eq:NLS} conserves the integrated Hamiltonian density, 
\[
H(z) = \int_{-\infty}^{\infty}{\mathcal{H}(z,t)\ud t}
\]
with $\mathcal{H}(z,t) \equiv \mathcal{H}_K + \mathcal{H}_{N\!L}$, $\mathcal{H}_K \equiv |u_t|^2/2$ and $\mathcal{H}_{N\!L} \equiv -|u|^4/2$. 

We split $H(z)$ accordingly into two terms $H(z) = H_K(z) + H_{N\!L}(z)$, where $H_K(z)$ represents the kinetic and $H_{N\!L}(z)$ the nonlinear (interaction) part. 

In the presence of higher-order dispersion, we have to include higher-order terms in $H_K$, but we verified that this does not change qualitatively the conclusions reported in Fig.~\ref{fig:hamiltonian}.

\subsection{Simulation of light propagation in an optical fibre}

We consider the following generalized NLS equation (GNLS)
\begin{multline}
	i\frac{\partial{U}}{\partial Z} + \sum_{k\ge 2}{\frac{i^k}{k!}\beta_k\frac{\partial^k U}{\partial T^k}}\\ +\gamma\left(1+i\tau_\mathrm{shock}\frac{\partial}{\partial T}\right)U\int_0^\infty{R(T')|U(T-T')|^2\ud T'}=0
	\label{eq:GNLS}
\end{multline}
with $R(T)=(1-f_R)\delta(T) + f_R h_R(T)$ is the nonlinear response function, which includes Kerr and Raman components, $Z$ and $T$ are dimensional propagation distance and time in a frame moving at the group velocity of the input pulse, $\beta_k$ are the dispersion coefficients, $\gamma$ is the nonlinear parameter and $\tau_\mathrm{shock}$ represents the coefficient of first order approximation of the nonlinearity dispersion \cite{Agrawal2012}.

Eq.~\eqref{eq:GNLS} is solved with by setting the following values: a pulse at wavelength $\lambda_0 = 1,064\,\mathrm{nm}$ of duration $T_\mathrm{FWHM} = 5\,\mathrm{ps}$ and peak power of $P=100$ W is injected in an optical fibre with nonlinear coefficient $\gamma = 15\, \mathrm{(W\, km)}^{-1}$ and the chromatic dispersion is expanded up to  $\beta_2 = -0.41\, \mathrm{ps^2/km}$, $\beta_3 = 0.0687 \,\mathrm{ps^3/km}$; we consider a fibre length of 25-40 m. Finally $10^{3}$ iterations with different realisations of noise (one photon per frequency bin with random phase).

Finally the time scale is chosen according to
$T_0 = T_{MI}\equiv2\pi/\Omega_{MI}$, with $\Omega_{MI}\equiv \sqrt{\frac{2\gamma P}{|\beta_2|}}$, corresponds to a rescaling of distance, $L_d = T_0^2/|\beta_2|$, amplitude  $U_0 = [\gamma P]^{-1/2}$
	and Hamiltonian $\tilde{H} = H/H_0$, with $H_0 \equiv T_0^2U_0^2/L_d$. This is used in Fig.~\ref{fig:collisions} and \ref{fig:hamiltonian} in order to compare the curves to the results of the WIM.

\subsection{A note on Raman acceleration}	
With reference to Fig.~\ref{fig:hamiltonian}(a), for a soliton with zero transverse velocity, the expected dependence of the kinetic part of the Hamiltonian is $H_K = H_K^0 + b P_0 Z^2$, where $H_K^0 = \frac{1}{2} \int_{- \infty}^{\infty}{|U_T|^2\ud T}$ is the kinetic part of the NLS Hamiltonian, $P_0= \int_{- \infty}^{\infty}{|U|^2\ud T}$ is the conserved total intensity and $b = \frac{4}{15}T_R A^4$ where $T_R = 3$ fs and $A$ is the amplitude of the soliton represents the effect of Raman acceleration. Instead we observe a jump at each collision and a subsequent steady linear increment of $H_K$, on account of friction-like effects due to inelastic collisions.

\bibliography{Roguesolitons2}

\end{document}